\documentclass[aps,physrev,preprint,superscriptaddress]{revtex4-2}
\usepackage{graphicx}% Include figure files
\usepackage{amsmath}

\begin{document}

%\preprint{}

\title{Shock wave transmission across solid–water interfaces in underwater electrical foil explosions}
% Shock wave transmission across solid materials in underwater electrical foil explosions


\author{Francesc Hernández Garcia}
\email[]{fhg@kth.se}
\author{Brian Bertacco}
\author{Tobia De Benedictis}
\author{Sebastián Rojas Mata}
\affiliation{ 
FLOW, Department of Engineering Mechanics, KTH Royal Institute of Technology, Stockholm, 100 44, Sweden %\\This line break forced with \textbackslash\textbackslash
}%
\author{Jergus Strucka}
\affiliation{ 
European X-Ray Free-Electron Laser Facility, Holzkoppel 4, 22869 Schenefeld, Germany}%
%\thanks{}
\author{Michael Liverts}%
\affiliation{ 
FLOW, Department of Engineering Mechanics, KTH Royal Institute of Technology, Stockholm, 100 44, Sweden %\\This line break forced with \textbackslash\textbackslash
}%
\homepage[]{https://www.flow.kth.se/}


\date{\today}

\begin{abstract}
The transmission of shock waves generated by underwater electrical foil explosions across solid materials with varying acoustic impedances and thicknesses is investigated. Copper foils with thicknesses of $10$ and $18~\mathrm{\mu m}$ are exploded on planar substrates of aluminum, copper, and 3D-printed PLA with thicknesses up to $5~\mathrm{mm}$ using a multi-gigawatt pulsed-power electrical discharge. The resulting shocks are visualized in water using high-speed shadowgraphy at $5$ million frames per second, while high-resolution fiber optic hydrophone measurements provide direct quantitative measurements of shock strength after transmission through the impedance-mismatched materials. One-dimensional magnetohydrodynamic simulations are used to reproduce the foil explosion dynamics and provide insight into the thermodynamic, acoustic, and electrical properties of the foil during the explosion. The results show that shock transmission is governed not only by the impedance mismatch at the solid--water interface but also by the finite thickness and dynamic response of the supporting material. These findings provide a quantitative framework for understanding and controlling shock transmission in underwater electrical foil explosions and establish the feasibility and limitations of using solid substrates as structural supports in complex shock-generating geometries.
\end{abstract}

%\keywords{}

\maketitle

\section{Introduction}

The electrical explosion of thin metal foils is a well-established technique for producing strong shock waves in materials~\cite{shelkovenko2020features}. When these explosions are generated underwater, gigapascal-scale pressures can be achieved behind the shocks~\cite{asmedianov2023underwater}. This process, known as underwater electrical foil explosion (UEFE), occurs when a fast, high-current pulse is discharged through a foil, rapidly heating it from the solid state into a dense, weakly ionized plasma that expands and drives a shock into water~\cite{grinenko2005strongly,asmedianov2026high}. UEFEs are relevant to a wide range of applications, including collision welding~\cite{vivek2013vaporizing}, microfabrication technologies~\cite{hasegawa2024uniformity}, dielectric flyer acceleration~\cite{asmedianov2025acceleration}, and nuclear fusion studies~\cite{nash2005design,slutz2016scaling}. Compared with the more widely studied underwater electrical wire explosion (UEWE) process~\cite{krasik2008underwater,hernandez2025characterization}, which generates cylindrical shocks, foil explosions can produce shocks that remain planar over a finite propagation distance. 

% Take this out in production please
\newpage 
The quasi-planarity of these shocks makes them useful for studies of supersonic multiphase flows, including Richtmyer--Meshkov hydrodynamic instabilities~\cite{strucka2023synchrotron,strucka2026passive}, water jets generated using V-shaped foils~\cite{asmedianov2026supersonic}, and shock--bubble interactions~\cite{bokman2025planar}. In such problems, the strong acoustic impedance mismatch between the medium in which the shock is generated, typically a liquid, and the medium into which it is transmitted, often a gas, plays a central role in the flow evolution. It is therefore important to quantify how transmission through acoustically mismatched interfaces affects the shock~\cite{hosseini2014shock}, particularly when the acoustic impedances of the involved media differ substantially. Simulations of multiphase problems driven by exploding conductors typically simplify the magnetohydrodynamic explosion process into an equivalent hydrodynamic formulation that reproduces the overall flow dynamics~\cite{maler2024multi,hernandez2026synchrotron,hernandez2026supersonic}. Although this approach has proven useful, it does not capture the actual pressure release associated with the explosion and the subsequent pressure evolution behind the shock front. Since direct pressure measurements in UEFEs remain scarce, the detailed explosion dynamics of the foil are often simplified. This motivates the present study, where high-resolution pressure measurements are carried out using a fiber optic hydrophone~\cite{staudenraus1993fibre,hernandez2025characterization}, providing direct experimental characterization of the shock pressure evolution in the near-field of the explosion and validation with magnetohydrodynamic simulations of the foil explosion process.

UEFEs are also relevant to shock focusing problems~\cite{apazidis2018shock}, where prescribed foil geometries enable the generation of converging shock waves with initial shock fronts that follow the foil shape. Previous work~\cite{rojas2026using} suggested that adding a $\sim 1.5~\mathrm{mm}$ thick plastic shell as a structural support, onto which the foil is pasted, can simplify the manufacturing of prescribed initial foil shapes. In this configuration, the foil explodes on the plastic shell, and the resulting shock propagates through it before being transmitted into water, where it focuses. The method appears to substantially improve the initial shape quality of the focusing shocks without significantly affecting their strength. This behavior could result from the acoustic impedance of the plastic shell being closer to that of water, although the effect of adding the shell was not quantified or fully understood. It remains an open question whether impedance-mismatched materials can be used as structural supports for exploding foils without significantly altering the generated shock.

In this work, experiments are conducted using a pulsed power generator (PPG)~\cite{akiyama2007industrial}, charged up to $23~\mathrm{kV}$, to electrically explode thin copper foils on a microsecond timescale. The foils are attached and exploded onto planar slabs of three target materials with different thicknesses, namely aluminum (Al), copper (Cu), and 3D-printed polylactic acid (PLA). These materials are selected to span a wide range of acoustic impedance mismatch at the material--water interface. The shocks are transmitted through the targets into the water, where the pressure of the transmitted shock is measured using a fiber optic hydrophone at a fixed stand-off distance from the explosion. The measured pressures are recovered using several common equations of state (EOS) for water to assess the sensitivity of the results to the EOS choice. The shock propagation in water after exiting the solid material is visualized using high-speed shadowgraphy at a rate of 5 million frames per second (Mfps). Experimental results from free-standing UEFEs are compared with a simplified one-dimensional magnetohydrodynamic simulation to reproduce the foil explosion process, validate the shock trajectory, and provide insight into thermodynamic, acoustic, and electrical Cu foil properties that are not accessible through direct experimental measurements. Together, these results provide new experimental measurements of the shock pressures generated by UEFEs, and establish how impedance-mismatched materials influence shock transmission, thereby assessing the feasibility of using these materials as supporting structural elements in UEFE experiments with complex geometries.

\section{Methods}

\subsection{Experiments}\label{Sec:Experiment}

\begin{figure}[h]
\centering
\includegraphics[width=0.6\linewidth]{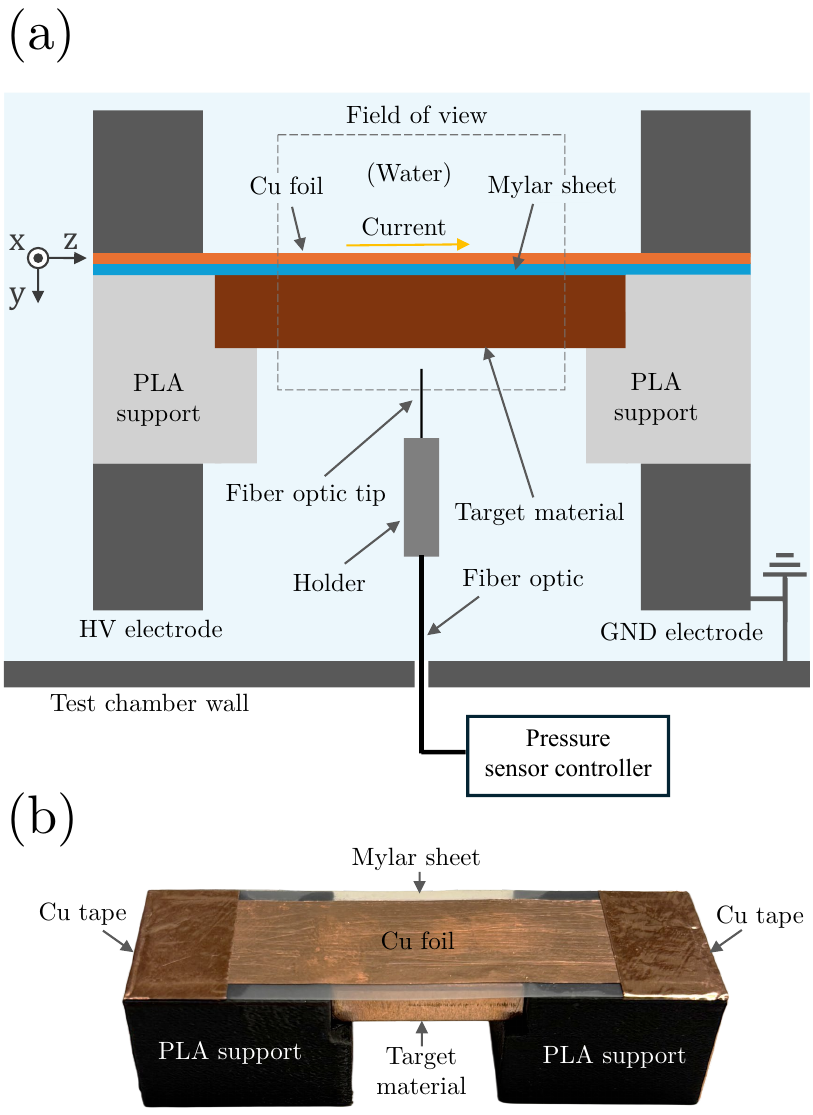}
\caption{Experimental configuration: (a) view along the laser beam axis; and (b) picture of a test sample.}
\label{fig:experimental_setup}
\end{figure}

The experimental setup consists of a PPG that delivers a microsecond-scale electrical discharge across thin Cu foils, causing them to explode and generate a shock in the surrounding medium. The PPG contains four capacitors connected in parallel, with a total capacitance of $10~\mathrm{\mu F}$, and has been used in previous studies~\cite{hernandez2025characterization,rojas2026using}. For all experiments reported in this work, the capacitors are charged to an initial voltage of $23~\mathrm{kV}$, corresponding to an available energy of $2.65~\mathrm{kJ}$. Figure~\ref{fig:experimental_setup}~(a) shows a diagram of the experimental configuration. The Cu foils are connected between the high-voltage (HV) and ground (GND) electrodes, which are separated by $40~\mathrm{mm}$. The HV electrode is connected to the capacitors, while the GND electrode is directly connected to the test chamber wall. The electrodes are $10~\mathrm{mm}$ thick stainless-steel cylinders with a central rectangular opening of $15~\mathrm{mm}$ height by $20~\mathrm{mm}$ width, into which the test sample is inserted and submerged in deionized water. The test sample consists of the exploding Cu foil, the target material, and an electrically insulating Mylar sheet placed between them, so that the discharge current flows only through the foil and not through the target material. This Mylar sheet is $0.1~\mathrm{mm}$ thick, $20~\mathrm{mm}$ wide, and $40~\mathrm{mm}$ long. Its thickness is chosen to be small to reduce shock attenuation, while still ensuring reliable electrical isolation between the exploding copper foil and the target materials. At this thickness, the ideal dielectric strength of Mylar in an underwater environment has been reported to be approximately $30~${--}$~40~\mathrm{kV}$ for DC voltage application \cite{inuishi1957electrical}. Although the present experiment uses a pulsed discharge, no dielectric breakdown is observed during the experiments, and all applied voltages remain below this reported level.

PLA supports are used to hold the sample and provide structural stiffness. These supports are press-fitted into the electrode holes to ensure good electrical connection. The region of the exploding Cu foil in contact with the electrodes is protected with $35~\mathrm{\mu m}$ Cu tape (which does not explode) to avoid damaging the foil at the electrode contact points. The exploding Cu foils are $40~\mathrm{mm}$ long, $14~\mathrm{mm}$ wide, and either $10$ or $18~\mathrm{\mu m}$ thick. This corresponds to available energy densities of $53$ and $29~\mathrm{kJ/g}$ for the $10$ and $18~\mathrm{\mu m}$ foils, respectively. The target material consists of PLA, Al, or Cu slabs with common thicknesses of $1.5$ and $5~\mathrm{mm}$, which are used to assess the effect of target material on the transmitted shocks. For Al, additional target thicknesses ranging from $0.05~\text{--}~1~\mathrm{mm}$ are tested to further study the dependence of shock strength on target thickness for a fixed material. All target materials are $20~\mathrm{mm}$ long and $15~\mathrm{mm}$ wide. Figure~\ref{fig:experimental_setup}~(b) shows an example picture of a sample consisting of an $18~\mathrm{\mu m}$ Cu foil placed on top of a $5~\mathrm{mm}$ thick Cu target before installation in the electrodes.

A high-speed camera is used to record 256 frames of the foil explosions at a rate of $5~\mathrm{Mfps}$. The image size is $400~\times 250$ pixels, corresponding to an effective pixel size of $\sim 70~\mathrm{\mu m/pixel}$ with our optical magnification. The frame exposure is $110~\mathrm{ns}$, of which $20~\mathrm{ns}$ are illuminated by a synchronized $500~\mathrm {W}$, $690~\mathrm{nm}$ pulsed laser. Shadowgraphy is used to visualize the shock after it transmits across the target material and propagates into the water. The characteristic blurring distance (i.e., shock smearing) during the exposure is smaller than the effective pixel size. The voltage at the HV electrode is measured using a high-voltage probe, while the discharge current is measured using a Rogowski coil~\cite{ward1993using}.

A fiber optic hydrophone pressure sensor~\cite{staudenraus1993fibre, hernandez2025characterization} is used to measure the pressure of the transmitted compression wave at a fixed stand-off distance of $6\pm 0.5~\mathrm{mm}$ from the exploding foil, in the direction of the target material. The sensor consists of a silica glass fiber tip with a diameter of $100~\mu m$ facing the target material, a plastic holder that keeps the fiber tip aligned in place, and a fiber optic cable connected to the sensor controller. A $690~\mathrm{nm}$ laser light is directed through the fiber and is partially reflected at the tip. Time-dependent changes in the refractive index of the medium, caused by changes in the local density, modify the reflected light signal. The corresponding pressure changes depend on the choice of EOS used to relate density and pressure in water; details on the pressure recovery procedure are provided in Sec.~\ref{Sec:pressMeasurements}. The constant measurement stand-off distance of $6~\mathrm{mm}$ is chosen to be larger than the maximum tested target thickness of $5~\mathrm{mm}$. This distance is also sufficiently large to prevent reflected light from the fiber laser on the target material surface from affecting the sensor calibration. The fiber tip is damaged after each explosion and therefore needs to be re-cleaved and calibrated in situ before each experiment. The pressure sensor output signal of $150~\mathrm{MHz}$ is acquired using an oscilloscope at $2~\mathrm{GS/s}$ with a built-in $700~\mathrm{MHz}$ bandwidth-limiting filter to mitigate aliasing. Additional details on triggering, signaling, optics, and hardware are described in Ref.~\citenum{hernandez2025characterization}.

\subsection{Material characterization}\label{Sec:MatChar}

The materials tested in this work are selected to span a wide range of acoustic impedance compared to water. Table~\ref{tab:material_acoustic_properties} presents their ambient density $\rho_0$, speed of sound $a_0$, compressibility $\beta_0$, acoustic impedance $Z$, and relative acoustic impedance with respect to water $Z/Z_{\mathrm{water}}$. The density of each material is measured by weighing the samples, while the sound speeds are taken from Refs.~\cite{selfridge1985approximate,marsh1980lasl,parker2010longitudinal}. The measured density of Mylar is slightly lower than typical reported values~\cite{selfridge1985approximate}. Cu and Al (1000 series in the H18 temper) are commercially pure materials, and their measured densities agree with nominal elemental bulk values. The PLA samples are fabricated by 3D printing commercial PLA filament with a Bambu Lab X1E printer, using a $0.4~\mathrm{mm}$ nozzle and a $0.08~\mathrm{mm}$ layer-height profile. A $100\%$ infill with straight filament paths is used to minimize internal air voids. For the $1.5~\mathrm{mm}$ thick samples, the measured PLA density agrees well with the theoretical value, indicating that the material is close to pure PLA. In contrast, for the $5~\mathrm{mm}$ thick samples, the measured density is approximately $80\%$ of that of pure PLA, indicating the presence of approximately $20\%$ air cavities. This porous morphology is also confirmed by visual inspection of the sample interior. The speed of sound of porous PLA is estimated by assuming a weighted average between pure PLA and the ambient speed of sound of air, taken as $343~\mathrm{m/s}$. This estimate is approximate but provides a basis for qualitatively assessing how the 3D-printed internal structure can affect the results.

\begin{table}[h]
\centering
\caption{Summary of material properties.}
\label{tab:material_acoustic_properties}
\begin{tabular}{lccccc}
\hline
Material & $\rho_0~(\mathrm{kg/m^{3}})$ & $a_0~(\mathrm{m/s})$ & $\beta_0~(\mathrm{TPa^{-1}})$ & $Z~(\mathrm{MRayl})$ & $Z/Z_{\mathrm{water}}$ \\
\hline
Water        & $998$  & $1490$      & $451$  & $1.49$  & $1.00$ \\
Mylar        & $1280$ & $2540$      & $121$  & $3.25$  & $2.19$ \\
PLA          & $1240$ & $2270$      & $156$  & $2.81$  & $1.89$ \\
PLA $(80\%)$ & $990$  & $\sim 1890$ & $283$  & $1.87$  & $1.26$ \\
Aluminium    & $2710$ & $6380$      & $9.1$  & $17.3$  & $11.6$ \\
Copper       & $8920$ & $4760$      & $4.9$  & $42.5$  & $28.6$ \\
\hline
\end{tabular}
\end{table}

The acoustic impedances are used to estimate the pressure reflection coefficient $\mathcal{R}$ and transmission coefficient $\mathcal{T}$ for a perturbation traveling from a medium with impedance $Z_1$ into a medium with impedance $Z_2$,

\begin{equation}
\mathcal{R} = \frac{Z_2-Z_1}{Z_2+Z_1},
\qquad
\mathcal{T} = \frac{2Z_2}{Z_1+Z_2}.
\end{equation}

The corresponding coefficients at the material--water interface are given in Table~\ref{tab:reflection_transmission_coefficients}. These coefficients quantify the reflected and transmitted pressure-wave amplitudes relative to the incident amplitude, with positive values indicating compression and negative values indicating rarefaction. This approach is valid for small-amplitude waves, that is, in the linear acoustic limit where the pressure wave Mach number is $M\sim 1$. For stronger shocks, the exact method to compute these coefficients is the impedance matching technique~\cite{forbes2012shock}. Preliminary analysis~\cite{bertacco2026experimental} shows that, compared with the linear approximation, using this technique changes the coefficients by less than $5\%$ in all cases (for the range of Mach numbers and materials studied in this work). Further discussion of how these coefficients affect the results is provided in Sec.~\ref{Sec:diffMaterials}.

\begin{table}[h]
\centering
\caption{Linear acoustic reflection and transmission coefficients at the material--water interface.}
\label{tab:reflection_transmission_coefficients}
\begin{tabular}{lcc}
\hline
Interface & $\mathcal{R}$ & $\mathcal{T}$ \\
\hline
Mylar--Water                  & $-0.37$ & $0.63$ \\
PLA--Water                    & $-0.31$ & $0.69$ \\
PLA $(80\%)$--Water & $-0.11$ & $0.89$ \\
Al--Water                     & $-0.84$ & $0.16$ \\
Cu--Water                     & $-0.93$ & $0.07$ \\
\hline
\end{tabular}
\end{table}

\subsection{Numerical simulations} \label{Sec:NumSim}

We simulate the UEFEs by adapting the model introduced in Hernández Garcia et al.~\cite{hernandez2025characterization}, which was originally developed and validated for UEWEs. The simulations aim to reproduce the dynamics of free-standing foil explosions and the thermodynamic properties in the region behind the generated shocks, without including any target material, providing a reference for evaluating the additional effects introduced by solid supports on shock transmission. The thermodynamic, acoustic, and electrical properties of the Cu foil during the explosion are analyzed without the effects introduced by interactions with surrounding materials. The numerical framework consists of three separate solvers: an electrical discharge circuit model, a zero-dimensional magnetohydrodynamic model for the Cu foil explosion, and a one-dimensional Euler solver for the surrounding water dynamics.

For the present work, the experimentally measured electrical waveforms are implemented directly in the simulations to reduce uncertainty associated with modeling the electrical discharge~\cite{revello2026hydrodynamic}, particularly from the triggering mechanism and unquantified sources of inductance. The time-dependent current $I(t)$ and total voltage $V(t)$ measured across the exploding foil are used as simulation inputs. The resistive voltage $V_R(t)$ across the foil is then computed as

\begin{equation}\label{Eq:resVolt}
    V_R(t) = V(t)-L(t)\frac{dI}{dt} - I(t)\frac{dL}{dt}.
\end{equation}

\noindent where $L(t)$ is the self-inductance of the foil. This quantity is estimated analytically~\cite{rosa1912formulas} using the vacuum permeability $\mu_0$, the foil length $l$, and the geometric mean distance $D(t)$ of the rectangular cross-section, with thickness $\tau(t)$ and width $w$, as

\begin{equation}\label{Eqimpedance}
    L(t) = \frac{\mu_0 l}{2\pi}
    \left[
        \ln\left(\frac{2l}{D(t)}\right) -1 + \frac{D(t)}{l}
    \right].
\end{equation}

\noindent This approach is justified since the voltage is measured directly across the foil, while additional inductive contributions from the circuit are minimized. Once $V_R(t)$ is known, the electric field $E_z(t)$ is calculated by assuming a uniform voltage drop along the foil,

\begin{equation}
    E_z(t) = \frac{V_R(t)}{l}.
\end{equation}

The electrical conductivity is evaluated as $\sigma(t)=\sigma(\rho_{\text{Cu}},T_{\text{Cu}})$ using a model based on the work by Chung et al.~\cite{chung2016numerical}, which builds on semiempirical work on UEWEs~\cite{stephens2014semiempirical}. The time evolution of the Cu foil density $\rho_{_{\text{Cu}}}$ and temperature $T_{_{\text{Cu}}}$ is then described by the mass and energy conservation equations, respectively,

\begin{gather}
    \frac{d\rho_{_{\text{Cu}}}}{dt} + \rho_{_{\text{Cu}}} \frac{\dot{\tau}}{\tau} = 0\;, \label{mhd11}\\
    \left( \frac{\partial \varepsilon_{_{\text{Cu}}}}{\partial T_{_{\text{Cu}}}} \right)_{\rho_{_{\text{Cu}}}} \frac{dT_{_{\text{Cu}}}}{dt} =  \left[ \frac{p_{_{\text{Cu}}}}{\rho_{_{\text{Cu}}}^2} - \left( \frac{\partial \varepsilon_{_{\text{Cu}}}}{\partial \rho_{_{\text{Cu}}}}\right)_{T_{_{\text{Cu}}}}  \right] \frac{d \rho_{_{\text{Cu}}}}{d t}  + \frac{1}{\rho_{_{\text{Cu}}}} \left( \sigma E_z^2  - \dot{\text{Q}}_R   \right)\;.  \label{mhd22}
\end{gather}

The term $\dot{\text{Q}}_R$ denotes the bulk radiative power loss per unit volume, modeled as grey-body radiation with an effective emissivity of 0.02~\cite{shi2021electrical}. An overdot denotes differentiation with respect to time. The quantities $\varepsilon_{_{\text{Cu}}}(t)$ and $p_{_{\text{Cu}}}(t)$ are the Cu specific internal energy and pressure, respectively. These are computed using the SESAME 3336 Cu EOS~\cite{sesame3336}, as $\varepsilon_{_{\text{Cu}}}(t) = \varepsilon_{_{\text{Cu}}}(\rho_{_{\text{Cu}}},\;T_{_{\text{Cu}}})$ and $p_{_{\text{Cu}}}(t) = p_{_{\text{Cu}}}(\rho_{_{\text{Cu}}},\;T_{_{\text{Cu}}})$. In the implementation, the foil is not allowed to expand in the solid phase while it heats up by enforcing $\dot{\tau}=0$. Upon reaching the melting temperature, the temperature is maintained constant until sufficient latent heat from Joule heating is supplied to complete the solid-to-liquid phase transition. After melting, $\dot{\tau}$ is allowed to evolve according to Eq.~(\ref{mhd11}). No analogous constraint is applied for the liquid-to-gas transition, as enforcing a constant-temperature condition in this regime led to numerical instabilities. This approach simplifies the phase transition physics while maintaining a consistent hydrodynamic response in water~\cite{hernandez2025characterization}.

The magnetic field at the foil surface is then estimated by treating the foil as a continuum of parallel line currents distributed uniformly across its rectangular cross-section. This approximation is justified in the thickness direction, since the skin depth during the discharge~\cite{keller2023design} is approximately $150~\mathrm{\mu m}$, significantly larger than the foil thicknesses considered here. Across the millimeter-scale foil width, however, the lateral skin effect may produce a nonuniform current distribution, concentrating current toward the edges~\cite{asmedianov2026high}. A differential current element $dI$ is given by $dI=\sigma E_z\,dy\,dx$, where $x$ and $y$ are the coordinates across the foil width and thickness, respectively. At the foil surface, the Biot--Savart law~\cite{griffiths2017introduction} leads to

\begin{equation}
    dB_x =
    \frac{\mu_0 \sigma E_z}{2\pi}
    \frac{\tau/2-y}{x^2+(\tau/2-y)^2}
    \,dx\,dy .
\end{equation}

Here, $dB_x$ is the differential contribution to the magnetic field component tangential to the foil surface and parallel to its width. Integrating over the rectangular cross-section, $-w/2\le x\le w/2$ and $-\tau/2\le y\le \tau/2$, gives

\begin{equation}
    B_x\!\left(\frac{\tau}{2}\right)
    =
    \frac{\mu_0 \sigma E_z \tau}{\pi}
    \tan^{-1}\left(\frac{w}{\tau}\right).
\end{equation}

The magnetic pressure at the foil surface is then computed as $p_B = B_x^2 /(2\mu_0)$. The pressure boundary condition at the contact surface (CS) between the Cu foil and water is given by $p_{\text{CS}} = p_{\text{Cu}} - p_B$. The water dynamics are modeled with the one-dimensional Euler equations in Cartesian coordinates,

\begin{equation} \label{EulerEq}
\frac{\partial \textbf{U}}{\partial t} +
\frac{\partial \textbf{F}\left(\textbf{U}\right)}{\partial x} = 0,
\end{equation}

\noindent where $\textbf{U}$ is the vector of conservative variables and $\textbf{F}\left(\textbf{U}\right)$ is the flux vector, respectively,

\begin{equation}
\textbf{U} =
\begin{bmatrix}
\rho \\
\rho u \\
\rho E
\end{bmatrix},
\quad 
\textbf{F}\left(\textbf{U}\right) =
\begin{bmatrix}
\rho u \\
\rho u^2 + p \\
\rho E u + p u
\end{bmatrix}.
\end{equation}

\noindent Here, $\rho$ is the density, $u$ is the velocity, $E$ is the specific total energy, and $p$ is the pressure. The system is closed using SESAME 7154 as the EOS~\cite{lyon1978sesame}, and a modified artificial upstream flux vector splitting scheme is used for the flux calculation~\cite{sun2003artificially}. Temporal integration is carried out with a fourth-order Runge-Kutta Dormand--Prince method~\cite{hairer1987solving}, and the spatial derivatives of the fluxes are computed using a fourth-order cell-centered finite difference scheme with a five-point stencil. The Courant--Friedrichs--Lewy (CFL) number is on the order of $10^{-2}$ due to the stability requirements of the coupled solvers, which share the same time step. A grid element size of $\sim 10~\mathrm{\mu m/element}$ is used to ensure full convergence of the results.

\section{Results and Discussion}

\subsection{Foil explosions and shock wave characterization}

\subsubsection{Shadowgraph visualization} \label{shadowgraph}

\begin{figure}[h]
\centering
\includegraphics[width=1\linewidth]{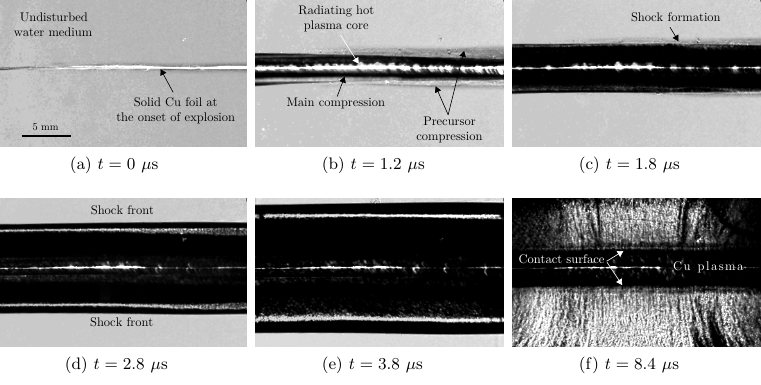}
\caption{Selected frames (a)--(f) from the explosion of a free-standing $18~\mathrm{\mu m}$ thick Cu foil in water, with the background subtracted. Time is relative to the explosion time $\pm 0.1~\mu\textbf{s}$.}
\label{fig:free_foil}
\end{figure}

We first perform experiments with free-standing foils, that is, without any target material through which the shock propagates. The relevant electrical waveforms driving these experiments are given in App.~\ref{Sec:Electrical}. Selected frames from an $18~\mathrm{\mu m}$ foil explosion are shown in Fig.~\ref{fig:free_foil}, imaged from the $x$-direction in Fig.~\ref{fig:experimental_setup}~(a), while the $10~\mathrm{\mu m}$ case is qualitatively analogous.

Figure~\ref{fig:free_foil}~(a) corresponds to the recorded frame closest to the explosion time, which is visually identified by the light emission from the foil as it undergoes different phase changes. In Fig.~\ref{fig:free_foil}~(b), we observe the appearance of a weak precursor compression, followed by a much stronger main compression. These features have previously been observed in Refs.~\cite{asmedianov2023underwater,rojas2026using}, in which the precursor compression was attributed to the solid-to-liquid phase change and the main compression to the liquid-to-gas phase change. The precursor compression is more visible for the $18~\mathrm{\mu m}$ foil than for the $10~\mathrm{\mu m}$ case. Since the thicker foil has a larger cross-sectional area, the current density evolution differs between the two cases, which affects the rate of energy deposition into the foil and delays vaporization after melting~\cite{sedoi1999current}. This allows the precursor compression to propagate farther before the main compression catches up. This process, referred to as shock formation, is shown in Fig.~\ref{fig:free_foil}~(c), and is captured in the simulations presented in Sec.~\ref{Sec:SimResults}. Figures~\ref{fig:free_foil}~(d)--(f) show the subsequent shock front propagation and the expansion of the hot Cu plasma, with the shock front remaining planar along the direction of the foil length.

\begin{figure}[h]
\centering
\includegraphics[width=1\linewidth]{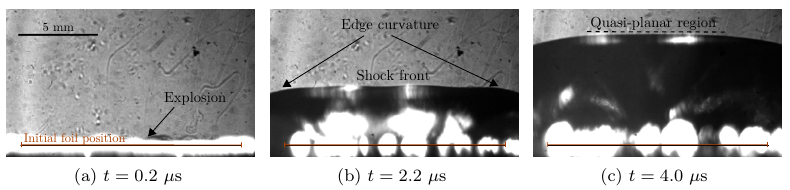}
\caption{Side view selected frames (a)--(c) from the explosion of a free-standing $10~\mathrm{\mu m}$ thick Cu foil in water. Time is relative to the explosion time $\pm 0.1~\mu\textbf{s}$.}
\label{fig:explosion_side_view}
\end{figure}

To assess the planarity of the shock front along the foil width direction, the experiment is repeated with the foils and electrodes rotated by $90^\circ$. Shadowgraph images of the explosion for a free-standing $10~\mathrm{\mu m}$ foil are shown in Fig.~\ref{fig:explosion_side_view}, imaged from the $z$-direction in Fig.~\ref{fig:experimental_setup}~(a), while the $18~\mathrm{\mu m}$ case is qualitatively analogous. Over time, the shock front's edges curve and start losing their two-dimensionality, while the central part remains quasi-planar. At a vertical distance of $6~\mathrm{mm}$ from the initial foil position, corresponding to the hydrophone measurement location, the quasi-planar region width is $\sim 8.5~\mathrm{mm}$ for the $10~\mathrm{\mu m}$ foil explosion and $\sim 7.5~\mathrm{mm}$ for the $18~\mathrm{\mu m}$ case. This relatively small difference may be attributed to the faster and more even $10~\mathrm{\mu m}$ foil explosion compared to the $18~\mathrm{\mu m}$ case. Although the thicker foil generates stronger shocks, the faster explosion of the thinner foil allows it to explode more uniformly and form the shock more rapidly. For the purposes of this study, these results indicate that shocks remain planar at the sensor position.

\subsubsection{Simulated shock dynamics} \label{Sec:SimResults}

\begin{figure*}[h]
    \centering
    \includegraphics[width=0.6\linewidth]{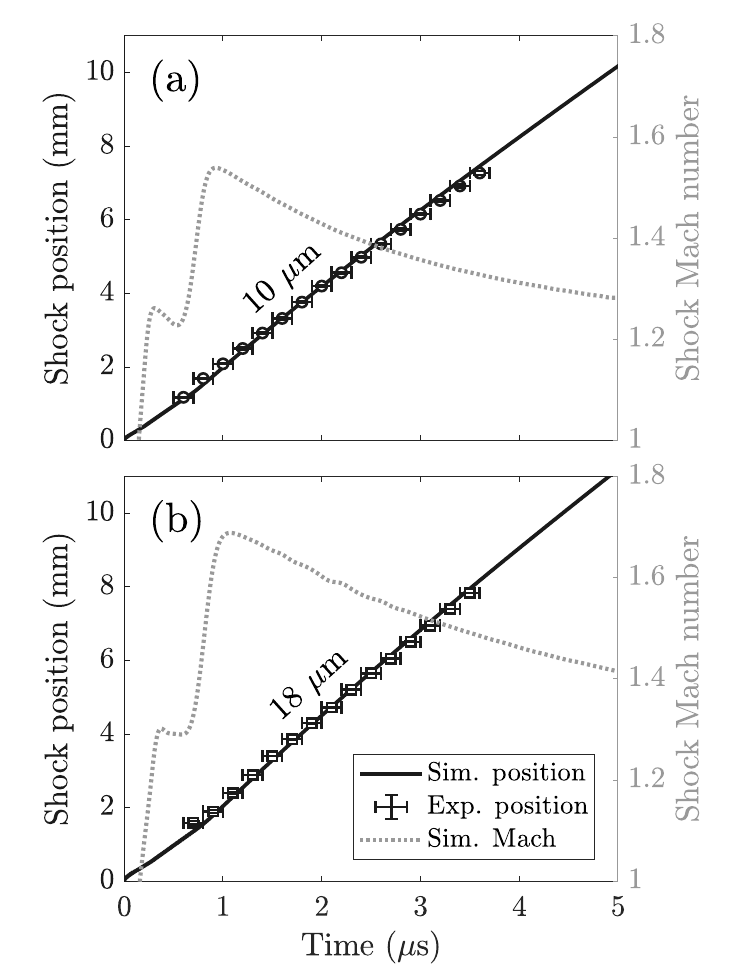}
    \caption{Experimental (Exp.) and simulated (Sim.) shock front positions, together with the corresponding simulated Mach number, for $10~\mathrm{\mu m}$ (a) and $18~\mathrm{\mu m}$ (b) foil explosions. Time is relative to the explosion time.}
    \label{fig:trajectory_no_mylar}
\end{figure*}

We now simulate the free-standing foil explosions using the numerical methods described in Sec.~\ref{Sec:NumSim}. We focus this section on the shock dynamics, and the reader is referred to Sec.~\ref{Sec:FoilProperties} for further discussion on the thermodynamic, electrical, and acoustic properties of the Cu plasma. The shock trajectories and Mach numbers are shown in Fig.~\ref{fig:trajectory_no_mylar}. The simulated and experimental trajectories agree within measurement uncertainty. The Mach number evolution shows two local maxima for both foil thicknesses within the first microsecond of the explosion. The first maximum is associated with the precursor compression generated during the solid-to-liquid phase change. The second maximum occurs when the main compression associated with the liquid-to-gas phase change catches up with this precursor compression, forming a single shock front. Shock formation is completed after propagation distances of $\sim 1.8$ and $\sim 2.1~\mathrm{mm}$ for the $10$ and $18~\mathrm{\mu m}$ foil explosions, respectively. The corresponding Mach numbers at these positions are 1.54 and 1.69, after which the shock front begins to decelerate.

\begin{figure*}[h]
    \centering
    \includegraphics[width=1\linewidth]{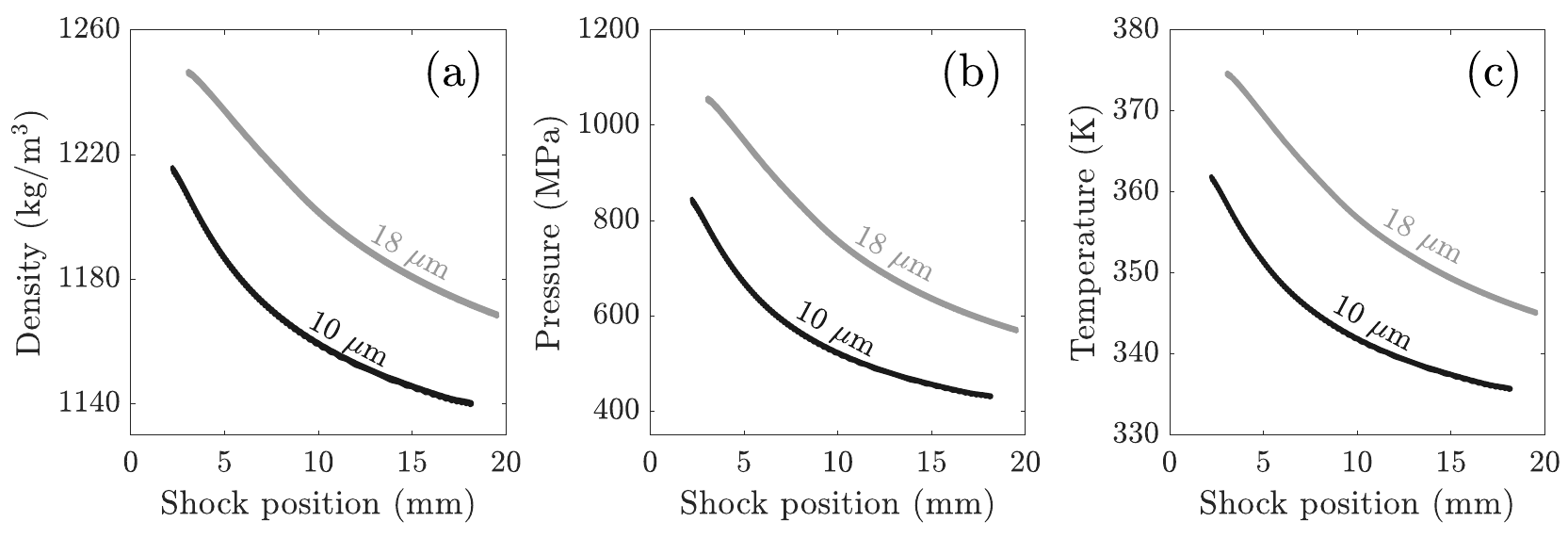}
    \caption{Simulated density (a), pressure (b), and temperature (c) behind the shock fronts.}
    \label{fig:sim_thermo_conditions}
\end{figure*}

Figure~\ref{fig:sim_thermo_conditions} shows the thermodynamic properties behind the shock front as a function of shock position, using SESAME 7154 as the EOS. All properties exhibit monotonically decreasing behavior, reflecting the decay of the shock strength. The $18~\mathrm{\mu m}$ foil consistently produces more extreme conditions behind the shock front. The spatial decay of these quantities can be well described by functional-form fits based on offset power-law exponential functions. At a distance of $6~\mathrm{mm}$ from the explosion, the post-shock density, pressure, and temperature are approximately $1180~\mathrm{kg/m^3}$, $630~\mathrm{MPa}$, and $349~\mathrm{K}$ for the $10~\mathrm{\mu m}$ foil, and $1228~\mathrm{kg/m^3}$, $923~\mathrm{MPa}$, and $367~\mathrm{K}$ for the $18~\mathrm{\mu m}$ foil, respectively. 

\subsubsection{Simulated foil plasma properties}\label{Sec:FoilProperties}

\begin{figure*}[h]
    \centering
    \includegraphics[width=1\linewidth]{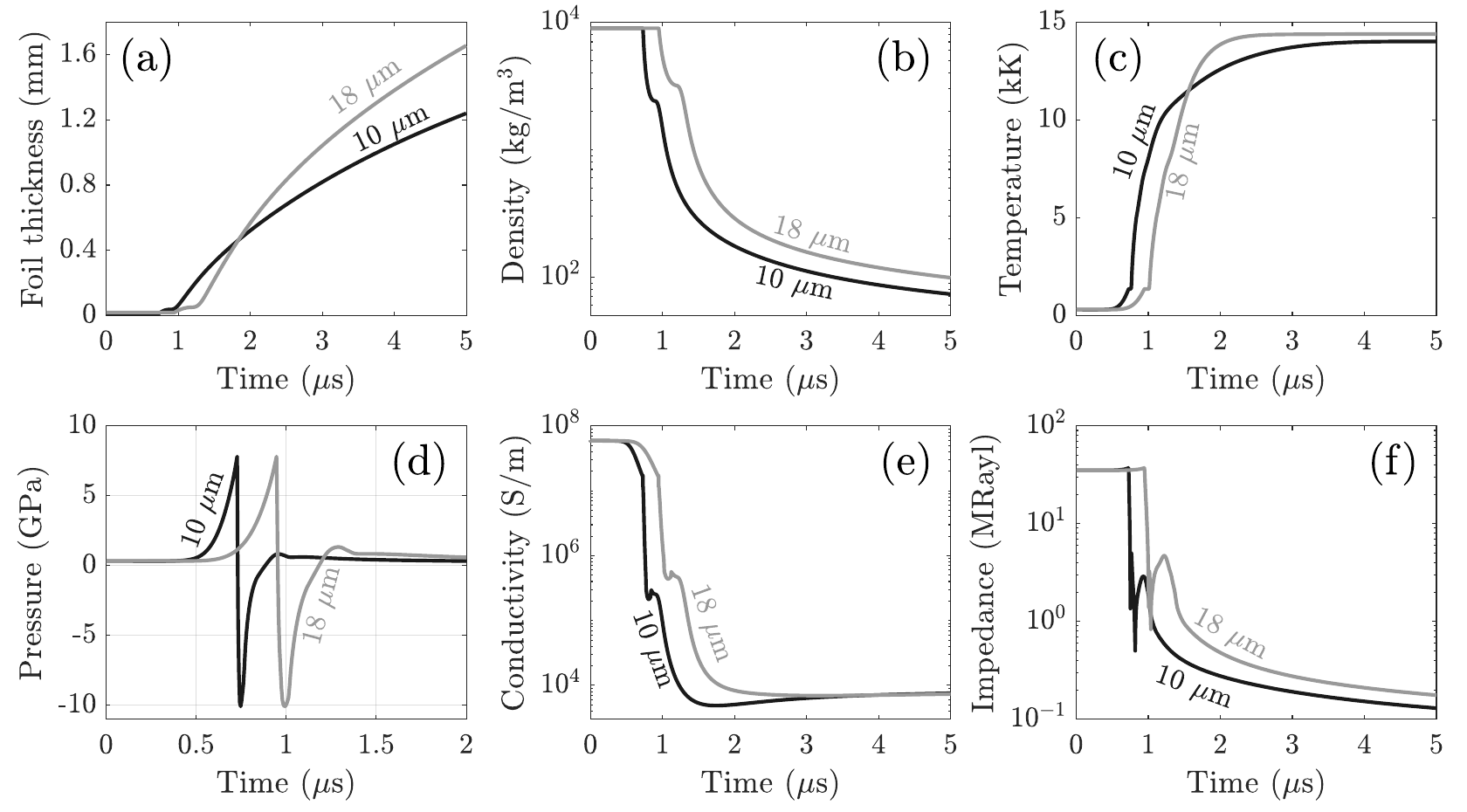}
    \caption{Simulated foil thickness (a), density (b), temperature (c), pressure (d), electrical conductivity (e), and acoustic impedance (f) for free-standing $10$ and $18~\mathrm{\mu m}$ Cu foil explosions. Time is relative to the current initiation.}
    \label{fig:MHD_properties}
\end{figure*}

The simulated thermodynamic, electrical, and acoustic properties of the expanding Cu plasma, obtained following the procedure described in Sec.~\ref{Sec:NumSim}, are shown in Fig.~\ref{fig:MHD_properties}. Figure~\ref{fig:MHD_properties}~(a) shows the foil thickness over time, referenced to current initiation. The model constrains the expansion of the foils in the solid phase. The $18~\mathrm{\mu m}$ foil starts expanding later than the $10~\mathrm{\mu m}$ foil, as its larger mass requires more time to reach the liquid state. Around $\sim 1~\mathrm{\mu s}$, both curves exhibit a concave-down elbow associated with the liquid-to-gas phase transition. The density, shown in Fig.~\ref{fig:MHD_properties}~(b), is held constant in the solid state until the transition to liquid, where a sudden density drop occurs at $0.73~\mathrm{\mu s}$ and $0.95~\mathrm{\mu s}$ after current initiation for the $10~\mathrm{\mu m}$ and $18~\mathrm{\mu m}$ foils. Respectively, the liquid-to-gas transition appears as a concave-up elbow at $0.95~\mathrm{\mu s}$ and $1.30~\mathrm{\mu s}$, indicating that the transition to the gas state occurs approximately $0.22~\mathrm{\mu s}$ and $0.35~\mathrm{\mu s}$ after liquefaction. Within the first microsecond of the explosion, the densities decrease to $\sim 2.4\%$ and $\sim 3.4\%$ of their initial values for the $10~\mathrm{\mu m}$ and $18~\mathrm{\mu m}$ foils. The density decreases monotonically, and the $18~\mathrm{\mu m}$ foil remains consistently denser than the $10~\mathrm{\mu m}$ foil throughout the simulation.

The temperature evolution is shown in Fig.~\ref{fig:MHD_properties}~(c). During the solid-to-liquid phase change, the temperature is imposed to remain constant. After liquefaction, the $18~\mathrm{\mu m}$ foil heats up faster, and both temperatures approach asymptotic values within the simulated time. This behavior reflects a balance between Joule heating, expansion work, and radiative losses, which may lead to a quasi-equilibrium asymptotic state~\cite{hernandez2025characterization}. The temperatures reached are $\sim 14.0~\mathrm{kK}$ and $\sim 14.4~\mathrm{kK}$ for the $10~\mathrm{\mu m}$ and $18~\mathrm{\mu m}$ foils, respectively. The pressure evolution is shown in Fig.~\ref{fig:MHD_properties}~(d). In the solid state, the pressure is allowed to increase, reaching a maximum value of $\sim 7.8~\mathrm{GPa}$ right before the solid-to-liquid transition. When the foil transitions to liquid and expansion begins, a sudden discontinuity appears, reducing the pressure to minimum values close to $-10.0~\mathrm{GPa}$. Such negative values may be non-physical, reflecting the extension of the employed EOS into the metastable regime in the absence of a Maxwell construction in the vapor dome~\cite{menikoff1989riemman,saurel2008modelling}, although they are not expected to significantly affect the overall dynamics, as the pressure rapidly recovers thereafter. Subsequently, the pressure rises to maximum values of $0.85$ and $1.35~\mathrm{GPa}$ for the $10~\mathrm{\mu m}$ and $18~\mathrm{\mu m}$ foils, respectively, at the liquid-to-gas transition time. After this maximum, the pressure decreases quasi-exponentially and remains positive for the rest of the simulation. The discontinuity at liquefaction results from suppressing foil expansion in the solid state, followed by sudden expansion and pressure release once the foil becomes liquid. The associated pressure drop acts as the disturbance that drives the precursor compression.

The electrical conductivity is given in Fig.~\ref{fig:MHD_properties}~(e), showing a decrease of $\sim 4$ orders of magnitude during the simulated foil expansion. For both foils, a small oscillating feature appears near $\sim 1~\mathrm{\mu s}$, corresponding to the time when the temperature is kept constant during the solid-to-liquid phase change. Finally, Fig.~\ref{fig:MHD_properties}~(f) presents the acoustic impedance of the Cu plasma, computed as the product of $\rho_{_\mathrm{Cu}}$ and the corresponding speed of sound, $a_{_\mathrm{Cu}}$. The speed of sound is calculated from the derivative of pressure with respect to density at constant entropy, $S$,

\begin{equation}
a_{_\mathrm{Cu}} =
\sqrt{
\left(
\frac{\partial p_{_\mathrm{Cu}}}{\partial \rho_{_\mathrm{Cu}}}
\right)_{S}
}.
\end{equation}

\noindent Since the pressure exhibits a discontinuity during the solid-to-liquid transition, this discontinuity is reflected in the calculated speed of sound and, consequently, in the acoustic impedance. The maximum impedance after liquefaction occurs at the liquid-to-gas transition time, coinciding with the positive local maximum in pressure. These maximum values are $2.90$ and $4.70~\mathrm{MRayl}$ for the $10~\mathrm{\mu m}$ and $18~\mathrm{\mu m}$ foils, respectively. These are still significantly larger than the impedance of water, but are roughly one order of magnitude smaller than that of solid Cu. At later times, the acoustic impedance decreases by another order of magnitude and becomes significantly smaller than the impedances of the test materials listed in Table~\ref{tab:material_acoustic_properties}. Therefore, a shock transmitted from the target materials to the plasma may initially experience an increasing impedance mismatch, which later evolves into a negative impedance mismatch over the experimental timescale. This may affect whether shock reverberations within the target materials are compressions or rarefactions, which is further discussed in Sec.~\ref{Sec:Reverberation}.

\subsubsection{Shock attenuation by the Mylar layer} \label{attMylar}

All experiments involving a target material include a $0.1~\mathrm{mm}$ Mylar sheet between the Cu foil and the target, which provides electrical insulation for the metal samples and is also used in the PLA shots for consistency. To quantify the influence of this layer, shock attenuation is evaluated at the hydrophone fiber tip position for the configuration with Mylar directly attached to the exploding Cu foil. This configuration is compared to the free-standing foil case, which serves as the reference for isolating the effect of the Mylar layer and assessing the additional effect of a target material. A quantitative comparison of the corresponding Mach numbers, including the simulated values, is given in Table~\ref{tab:mach_numbers}.

\begin{table}[h]
\centering
\caption{Numerical (Num.) and experimental (Exp.) Mach numbers measured at $6~\mathrm{mm}$ from the foil, with and without a $0.1~\mathrm{mm}$ Mylar sheet.}
\label{tab:mach_numbers}
\begin{tabular}{cccc}
\hline
Foil & Num. & Exp. & Exp. + Mylar $0.1~\mathrm{mm}$\\
\hline
$10~\mathrm{\mu m}$ & $1.34$ & $1.37 \pm 0.02$ & $1.27 \pm 0.01$ \\
$18~\mathrm{\mu m}$ & $1.52$ & $1.50 \pm 0.06$ & $1.43 \pm 0.01$ \\
\hline
\end{tabular}
\end{table}

The Mach number decreases from $1.37 \pm 0.02$ to $1.27 \pm 0.01$ for the $10~\mathrm{\mu m}$ foil, and from $1.50 \pm 0.06$ to $1.43 \pm 0.01$ for the $18~\mathrm{\mu m}$ foil when Mylar is included. To quantify the pressure attenuation associated with these changes, the pressure behind the shock is expressed as a function of Mach number, $p(M)$. This relation is EOS-dependent, and its derivation is provided in App.~\ref{Sec:HugoniotCurves} for several EOS. Here, SESAME 7154 is used to remain consistent with the numerical simulations. Using the $p(M)$ relation with the Mach numbers given in Table~\ref{tab:mach_numbers}, the effective pressure transmission between the cases with and without Mylar is $0.69 \pm 0.05$ for the $10~\mathrm{\mu m}$ foil and $0.83 \pm 0.13$ for the $18~\mathrm{\mu m}$ foil. These effective transmissions are larger than the transmission coefficient of $0.63$ estimated for the Mylar--water interface in Table~\ref{tab:reflection_transmission_coefficients}, possibly because the pressure of the expanding Cu plasma remains high after shock formation, as shown in Sec.~\ref{Sec:FoilProperties}. Therefore, the plasma continues to drive the Mylar from the foil side, and pressure perturbations from the plasma can continue to contribute to the shock transmitted into water because the pressure release is not instantaneous. Mylar bending and deformation within the timescale of the experiment may also contribute to the measured excess of shock strength by modifying the confinement of the expanding plasma.

A secondary effect is the finite-thickness response of the Mylar layer. After the initial shock reaches the Mylar--water interface, part of the wave reflects back through the Mylar and returns to the Mylar--plasma interface, producing a reverberation analogous to effects discussed in Sec.~\ref{Sec:Reverberation}. For a $0.1~\mathrm{mm}$ Mylar layer and a sound speed of $2540~\mathrm{m/s}$, this round trip takes approximately $80~\mathrm{ns}$. As shown in Sec.~\ref{Sec:FoilProperties}, the acoustic impedance of the Cu plasma decreases significantly over this timescale. At approximately $80~\mathrm{ns}$ after the liquid-to-gas phase transition, the simulated Cu plasma impedances are $\sim 1.50$ and $\sim 3.05~\mathrm{MRayl}$ for the $10$ and $18~\mu\mathrm{m}$ foils, respectively. Since the Mylar impedance is $3.25~\mathrm{MRayl}$, the reflected perturbation at the Mylar--plasma interface is expected to be a second rarefaction that decreases the pressure transmitted from the foil side. Nevertheless, the reverberation modifies the pressure history at the Mylar boundaries, while the sustained plasma pressure remains the dominant reason for the measured effective transmission being larger than the isolated Mylar--water interface value.

\subsubsection{Hydrophone pressure measurements} \label{Sec:pressMeasurements}

Pressure measurements are taken at a distance of $6 \pm 0.5~\mathrm{mm}$ from the exploding foils using the fiber optic hydrophone pressure sensor described in Sec.~\ref{Sec:Experiment}. This diagnostic provides high temporal and spatial resolution, but it relies on the detection of $690~\mathrm{nm}$ light reflected at the fiber tip. Therefore, external emission at this wavelength can corrupt the pressure signal. In our experiments, the Cu plasma acts as a light source, since it reaches temperatures of approximately $14~\mathrm{kK}$ according to the simulations shown in Sec.~\ref{Sec:FoilProperties}. Although the dominant emission is expected to lie in the ultraviolet, part of the emitted spectrum extends into the sensor wavelength. This light contamination is more severe when the sensor faces an exploding foil than in measurements of exploding wires~\cite{hernandez2025characterization}, because the light emission from wire explosions is azimuthally distributed, whereas the foil emission is directed toward the fiber tip. For this reason, it has not been possible to directly measure the shock pressure generated by a free-standing foil explosion with this method, as significant light contamination remains even by the time the shock reaches the fiber tip. However, the shock pressure transmitted through the Mylar can be measured by painting the Mylar surface with matte black paint, making it optically opaque at $690~\mathrm{nm}$ and preventing light from the foil explosion from reaching the fiber. The target materials are inherently opaque, so no additional painting is required. As a result, only the shocks transmitted through the painted Mylar or the target materials are measured.

\begin{figure*}[h]
    \centering
    \includegraphics[width=1\linewidth]{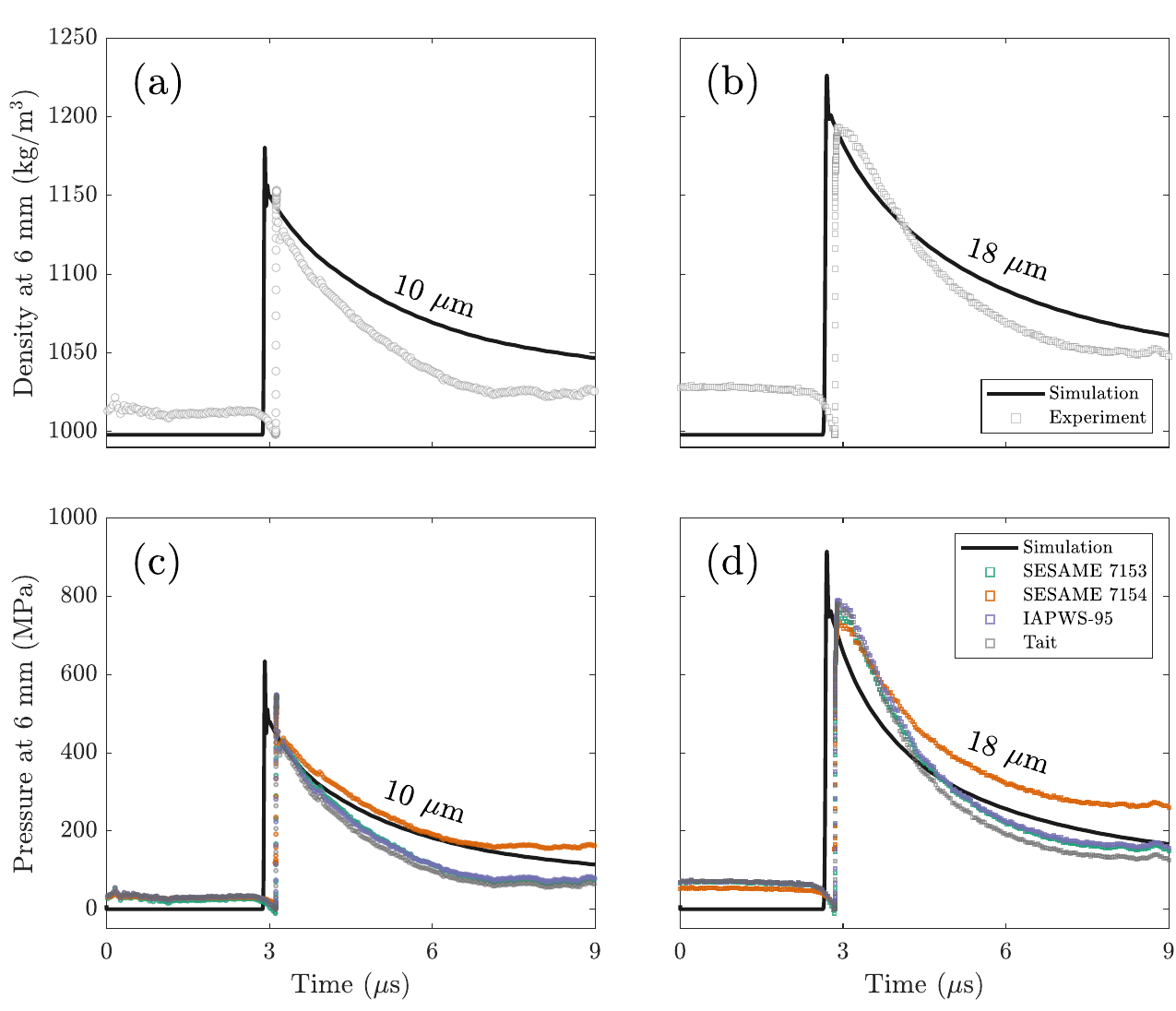}
    \caption{Density (a,b) and pressure (c,d) at a distance of $6~\mathrm{mm}$ from the foil explosions for the $10~\mathrm{\mu m}$ (a,c) and $18~\mathrm{\mu m}$ (b,d) foil explosions. The free-standing simulations are compared with fiber optic hydrophone measurements obtained after shock transmission through $0.1~\mathrm{mm}$ thick black-matte painted Mylar sheet. The retrieved pressure is shown for different equations of state. Time is relative to the explosion time $\pm 0.1~\mu\textbf{s}$.}
    \label{fig:pressure_sensor_EOS_curves}
\end{figure*}

Converting the density measurements into pressure requires an EOS. In this work, SESAME 7153, SESAME 7154~\cite{lyon1978sesame}, IAPWS-95~\cite{wagner2002iapws}, and Tait equation~\cite{ridah1988shock} are used to evaluate how the choice of water EOS affects the recovered pressure. Except for Tait, which is isothermal, these EOS also require temperature as input. This temperature is estimated from the Hugoniot curves derived in App.~\ref{Sec:HugoniotCurves} by identifying the shock state associated with the measured density. The pressure is then evaluated from each EOS at that state. In the absence of direct pressure measurements for the free-standing foil explosions, we use measurements of the shocks transmitted through the painted Mylar and compare them with numerical results, as given in Fig.~\ref{fig:pressure_sensor_EOS_curves}.

Density measurements are shown in Figs.~\ref{fig:pressure_sensor_EOS_curves}~(a,b). For the $10~\mathrm{\mu m}$ foil, the peak density is $1179~\mathrm{kg/m^3}$ in the simulation and $1153~\mathrm{kg/m^3}$ in the experiment. For the $18~\mathrm{\mu m}$ foil, the corresponding values are $1226$ and $1194~\mathrm{kg/m^3}$. The lower experimental densities are consistent with the shock strengths shown in Table~\ref{tab:mach_numbers}, since these measurements include the Mylar sheet, which the simulations ignore. After the shock passage, this difference also causes the simulated density to decay more slowly than the measured one, as propagation through the Mylar layer attenuates the post-shock profile more than water would. The corresponding pressures are shown in Figs.~\ref{fig:pressure_sensor_EOS_curves}~(c,d). Using SESAME 7154, the simulated peak pressures are $615~\mathrm{MPa}$ and $910~\mathrm{MPa}$ for the $10$ and $18~\mathrm{\mu m}$ foils, respectively. The experimental pressures vary between $520$--$550~\mathrm{MPa}$ for the $10~\mathrm{\mu m}$ foil and $740$--$800~\mathrm{MPa}$ for the $18~\mathrm{\mu m}$ foil, depending on the EOS used. This spread shows that the choice of EOS has a significant quantitative effect on the pressure recovered from hydrophone measurements. Among the EOS considered, SESAME 7154 gives the closest agreement with the simulated post-shock pressure profile, likely because the same EOS is used in the simulations.

A small density oscillation appears just behind the shock's peak in both the simulated and experimental waveforms, possibly due to ringing artifacts also observed in wire explosion measurements~\cite{hernandez2025characterization}. In the simulations, this may arise from the high-order treatment of the spatial derivatives of the flux vectors. In the measurements, it may be caused by the acquisition of the high-frequency shock front signal, which can produce post-front ringing after a sharp transient due to Gibbs phenomena associated with the spectral reconstruction of the signal~\cite{gottlieb1997gibbs}. Alternatively, the oscillation may have a physical origin. The shock front forms through the nonlinear interaction between the precursor compression and the main compression generated later in the explosion. As the main compression catches up with the precursor, the flow behind the front readjusts toward a single shock state, which can produce a small compression--rarefaction perturbation near the merged front. 

In addition to the possible origins of the observed oscillation, the pressure measurements are also affected by several experimental uncertainties. One source is fiber optic positioning, which results in a tip-to-foil distance in the range of $5.5~${--}$~6.5~\mathrm{mm}$. This corresponds to a density variation of approximately $5~${--}$~10~\mathrm{kg/m^3}$ according to the simulation results presented in Fig.~\ref{fig:sim_thermo_conditions}. This effect is expected to be smaller in experiments with solid targets, where attenuation within the material reduces sensitivity to probe position. Another source of uncertainty is possible overpressure at the fiber tip due to reflection of the shock at the $100~\mathrm{\mu m}$-thick tip cross-section, which may contribute up to $\sim 10\%$ of the measured excess density above ambient~\cite{hernandez2025characterization}. Additional unquantified sources of uncertainty, such as fiber misalignment, tip deformation, and light contamination, may also influence the results.

\subsection{Exploding foils on target materials}

\subsubsection{Shock transmission through different materials} \label{Sec:diffMaterials}

\begin{figure}[h]
    \centering
    \includegraphics[width=1\linewidth]{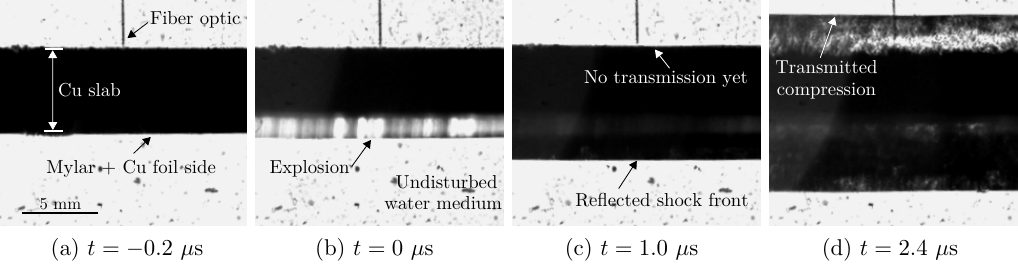}
    \caption{Selected frames (a)--(d) from the explosion of a $10~\mathrm{\mu m}$ thick Cu foil on a $5~\mathrm{mm}$ thick Cu slab. Time is relative to the explosion time $\pm 0.1~\mu\textbf{s}$.}
    \label{fig:slab_foil}
\end{figure}

Several target materials are placed on top of the Mylar to study the attenuation of the transmitted shocks at a fixed sensor position of $6\pm0.5~\mathrm{mm}$ from the explosion. The materials tested are PLA, Al, and Cu, with thicknesses of $1.5$ and $5.0~\mathrm{mm}$. Selected shadowgraph images of a $10~\mathrm{\mu m}$ thick Cu foil exploding on a $5~\mathrm{mm}$ thick Cu slab are given in Fig.~\ref{fig:slab_foil}. The configuration before the explosion is shown in Fig.~\ref{fig:slab_foil}~(a), with the fiber optic tip facing the Cu slab at a distance of $\sim 1~\mathrm{mm}$ from its top surface. The explosion time corresponds to the frame shown in Fig.~\ref{fig:slab_foil}~(b). Figures~\ref{fig:slab_foil}~(c,d) show the reflected shock after the foil explosion and the transmitted compression propagating above the sample. Whether this transmitted compression remains a shock or becomes an acoustic wave is assessed from the transmitted Mach number and the pressure sensor signal. 

\begin{figure}[h]
    \centering
    \includegraphics[width=1\linewidth]{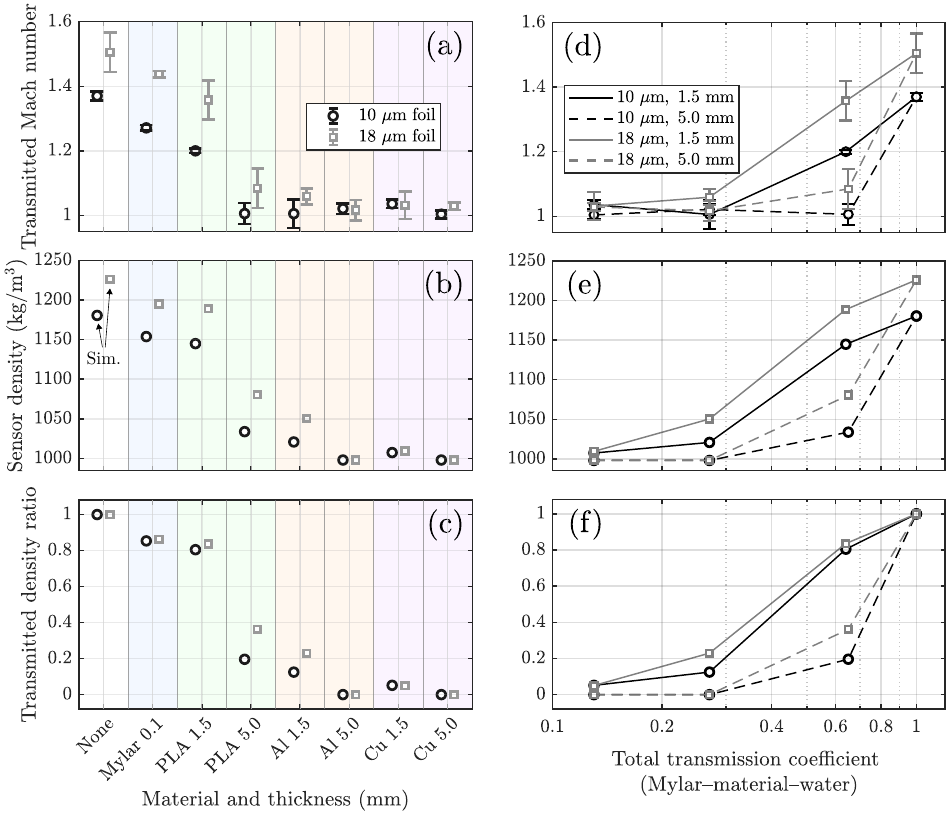}
    \caption{Mach number (a,d), measured sensor density (b,e), and density ratio (c,f) for shocks transmitted across Mylar, PLA, Al, and Cu of different thicknesses, measured at a distance of $6~\mathrm{mm}$ from the foil explosions. Panels (a)--(c) show the data for the individual materials, while panels (d)--(f) show the same data plotted as a function of the total transmission coefficient of the Mylar--material--water system, excluding the Mylar-only case. The legend in (d) indicates foil and sample thickness, respectively. The density ratio is defined as the measured density normalized by the simulated density for the case without material.}
    \label{fig:mat_comparison}
\end{figure}

These results are shown in Fig.~\ref{fig:mat_comparison}~(a)--(c) for all tested materials, including the measured Mach number, the maximum density at the sensor position, and the density ratio relative to the simulated free-standing case. Increasing the material thickness decreases the strength of the transmitted shock in all cases, as shown by the reduction in both Mach number and measured density. Both quantities are consistently higher for the $18~\mathrm{\mu m}$ foil than for the $10~\mathrm{\mu m}$ foil. For PLA, the $1.5~\mathrm{mm}$ thick sample decreases the density ratio by only $3$--$4\%$ relative to the case with only Mylar, indicating that thin PLA samples have a limited effect on the transmitted shock strength. Increasing the PLA thickness to $5~\mathrm{mm}$ decreases the density ratio from $0.80$ to $0.20$ for the $10~\mathrm{\mu m}$ foil and from $0.84$ to $0.36$ for the $18~\mathrm{\mu m}$ one, with $M<1.1$ in both cases. For Al and Cu, the shocks are strongly attenuated already for the $1.5~\mathrm{mm}$ thick samples, with measured Mach numbers of $M<1.06$ for Al and $M<1.04$ for Cu. The density ratios remain significantly large for Al, reaching $\sim 13\%$ and $\sim 23\%$ for the $10$ and $18~\mathrm{\mu m}$ foils, respectively, while they decrease to $\sim 5\%$ for Cu. Increasing the thickness to $5~\mathrm{mm}$ brings the measured Mach numbers to the sonic limit within measurement error, and the density measurement is not able to resolve the disturbance. This indicates that attenuation in the material is strong enough to reduce the transmitted shock to the weak limit of an acoustic wave. The data are plotted in Fig.~\ref{fig:mat_comparison}~(d)--(f) as a function of the total transmission coefficient of the Mylar--material--water system configuration, calculated as the product of the transmission coefficients at the Mylar--material and material--water interfaces. This representation reduces each material configuration to a single transmission coefficient, allowing the different materials to be compared on a common basis. In all cases, the transmitted shock strength consistently increases with the total transmission coefficient, for both sample and exploding foil thicknesses considered.

The observed trends can be interpreted from material attenuation inside the samples and shock transmission through the sample--water interface. Thicker materials reduce the measured Mach numbers more strongly, consistent with stronger shock attenuation inside the solid than in the liquid phase. For a fixed external driving pressure, heavier and less compressible media generate higher pressure decay rates~\cite{meyers1994shock}. A shock propagating through a longer path in a less compressible medium is therefore expected to lose more pressure than one propagating through a more compressible medium. Based on the values in Table~\ref{tab:material_acoustic_properties}, water is more compressible and lighter than all the materials tested here. This explains why thinner samples attenuate less, since the shock propagates through a longer path in water and a shorter path in the solid. For PLA $80\%$ (corresponding to the $5~\mathrm{mm}$ thick PLA sample), attenuation is expected to be enhanced by its internal porous structure and the presence of air voids. For a given thickness, PLA attenuates the least, followed by Al and then Cu. This trend is consistent with the transmission coefficients in Table~\ref{tab:reflection_transmission_coefficients}, where the Cu--water interface is one order of magnitude less transmissive than the PLA--water interface. All reflection coefficients are negative, meaning that the reflected pressure wave is a rarefaction~\cite{meyers1994shock}. This can affect the transmitted signal through reverberation inside the finite-thickness samples.

\subsubsection{Effect of varying aluminum target thickness}\label{Sec:targetThickness}

\begin{figure}[h]
    \centering
    \includegraphics[width=0.6\linewidth]{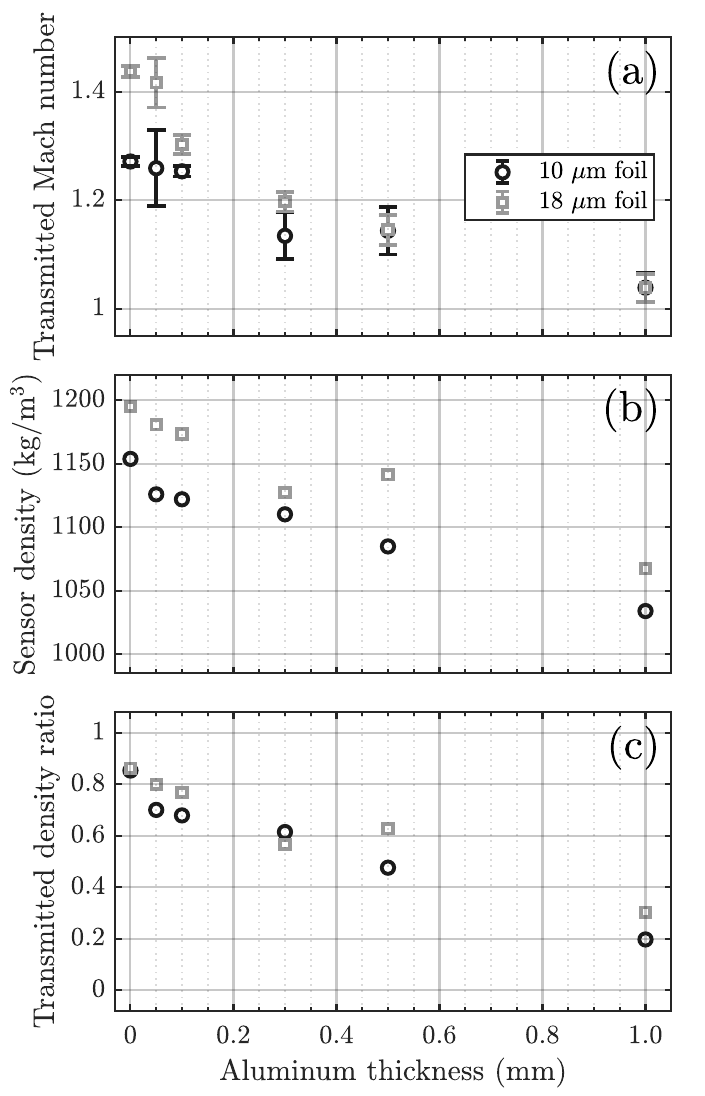}
    \caption{Mach number (a), measured sensor density (b), and density ratio (c) for shocks transmitted across Al samples of varying thickness up to $1~\mathrm{mm}$, measured at a distance of $6~\mathrm{mm}$ from the foil explosions. The $0~\mathrm{mm}$ case corresponds to the $0.1~\mathrm{mm}$ thick Mylar layer without an additional Al sample. The density ratio is defined as the measured density normalized by the simulated density for the case without material.}
    \label{fig:thin_foil_comparison}
\end{figure}

We now focus on Al samples with decreasing thicknesses of $1.0,\;0.5,\;0.3,\;0.1,$ and $0.05~\mathrm{mm}$. The results are shown in Fig.~\ref{fig:thin_foil_comparison}, including the transmitted Mach number, the maximum measured sensor density, and the density ratio with respect to the simulated free-standing case at a distance of $6~\mathrm{mm}$ from the explosion. The results show a progressive loss of shock strength as the Al thickness increases, consistent with stronger attenuation over a longer path in the solid. The $0.05~\mathrm{mm}$ thick sample decreases the density ratio by $5~${--}$~15\%$ relative to the case with only Mylar, whereas the $1.0~\mathrm{mm}$ thick sample increases this reduction to $55~${--}$~65\%$. For the thinner Al targets, the transmitted shock remains significantly stronger than expected from the Al--water transmission coefficient of $0.16$. This behavior is consistent with the observation that target samples with thicknesses $\leq 0.5~\mathrm{mm}$ undergo localized perforation and significant deformation within the experiment timescale. In contrast, the $1.0~\mathrm{mm}$ thick sample remains mostly intact with barely any visible deformation. This deformation can affect how effectively the post-shock pressure generated by the Cu plasma is transferred through the sample. If the material does not deform during the explosion, the plasma stays more confined, and the pressure transfer is expected to approach the ideal shock transmission through the material. If the material bends or perforates, the plasma can expand more freely, and the pressure release becomes more similar to that of the free-standing case. This shows that efficient shock attenuation requires not only sufficient material thickness, but also the ability of the material to maintain sufficient structural integrity over the timescale of the experiment.

\subsubsection{Shock wave reverberations}\label{Sec:Reverberation}

\begin{figure*}[h]
    \centering
    \includegraphics[width=1\linewidth]{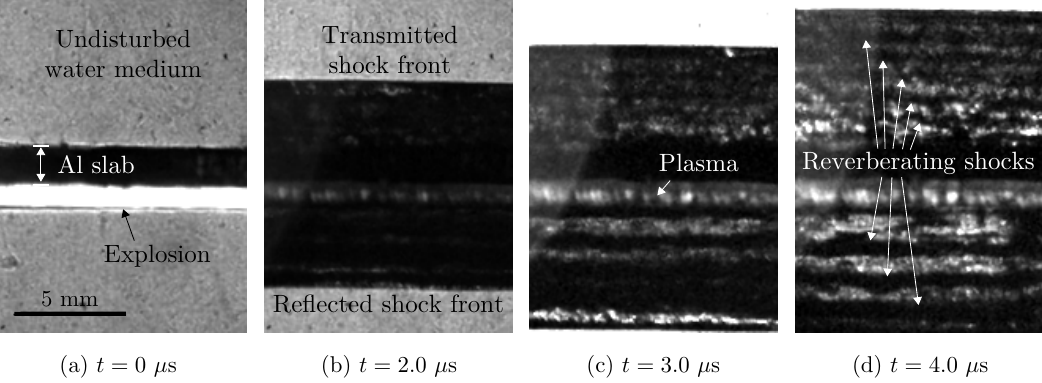}
    \caption{Selected frames (a)--(d) from the explosion of a $10~\mathrm{\mu m}$ thick Cu foil on a $1.5~\mathrm{mm}$ thick Al slab. The contrast in panels (c,d) is enhanced to improve the visualization of the reverberating shocks. Time is relative to the explosion time $\pm 0.1~\mu\textbf{s}$.}
    \label{fig:reverberation_video}
\end{figure*}

When a shock propagates inside a finite-thickness target, the wave transmitted into the sample can reflect at the material--water interface and travel back toward the material--plasma interface. Subsequent reflections at both boundaries can generate a sequence of pressure waves trapped inside the sample, with part of the disturbance transmitted into the surrounding media at each interaction. This process is commonly referred to as shock reverberation. It occurs when the acoustic impedance mismatch at the confining interfaces is large enough to produce significant reflections, which is widely used in dynamic compression experiments to infer material properties under extreme conditions~\cite{kondo1981multiple,knudson2003use,DeMange2009reverberation}, and has been previously reported in Al targets~\cite{plume2015systematic,gonzalez2018shock}. Shadowgraph images of the reverberations are shown in Fig.~\ref{fig:reverberation_video} for a $10~\mathrm{\mu m}$ thick Cu foil exploding on a $1.5~\mathrm{mm}$ thick Al sample.

\begin{figure*}[h]
    \centering
    \includegraphics[width=1\linewidth]{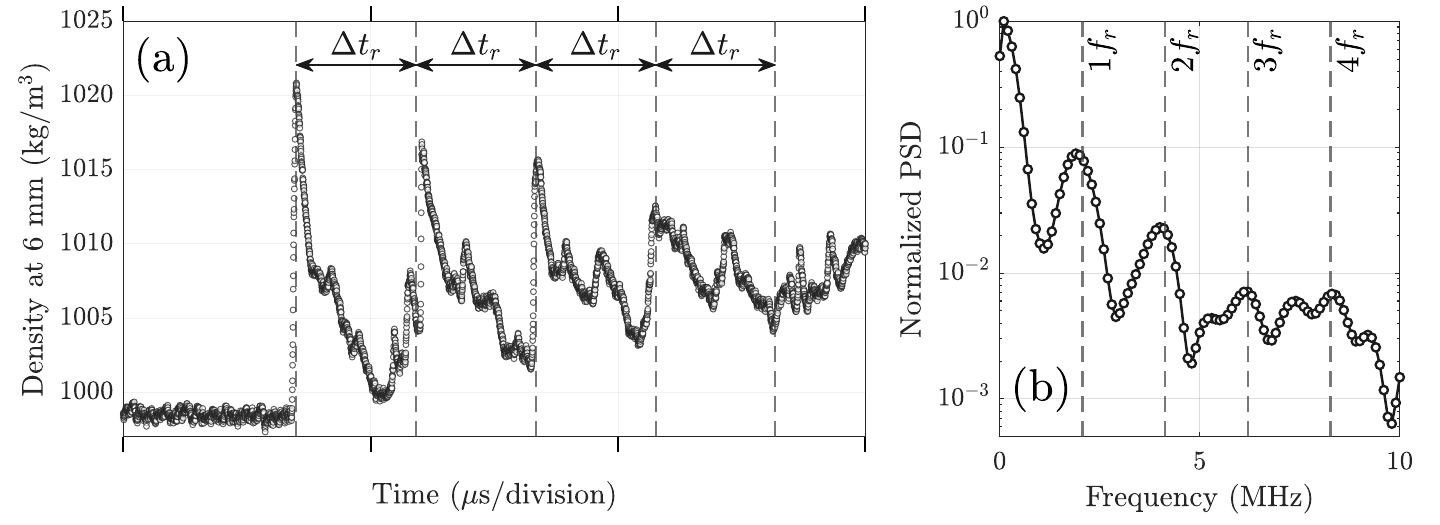}
    \caption{Measured density signal at a distance of $6~\mathrm{mm}$ from the explosion of a $10~\mathrm{\mu m}$ thick Cu foil on a $1.5~\mathrm{mm}$ thick Al slab (a), together with the corresponding power spectral density (PSD) (b).}
    \label{fig:reverberation_sensor}
\end{figure*}

Figures~\ref{fig:reverberation_video}~(c,d) show secondary fronts behind both the transmitted and reflected shocks, indicating that part of the pressure disturbance remains inside the Al sample after the first transmission and reflection events. Clear reverberation is observed only with the $1.0$ and $1.5~\mathrm{mm}$ thick Al samples, with a similar but weaker analogous signature also visible with the $1.5~\mathrm{mm}$ thick Cu sample. No clear reverberation is seen with PLA, possibly because the impedance mismatch with water is smaller and the transmitted disturbance is less strongly confined within the material. It is also not observed with the $5~\mathrm{mm}$ thick samples, where material attenuation likely weakens the internal wave before repeated reflections can be achieved, or the transmitted waves are too weak to be visualized in water. For the thinner Al samples discussed in Sec.~\ref{Sec:targetThickness}, local perforation and strong deformation are expected to disrupt the boundary confinement, preventing a stable reverberating wave from forming. The corresponding density signal for a $10~\mathrm{\mu m}$ thick Cu foil exploding on a $1.5~\mathrm{mm}$ thick Al sample, measured by the fiber optic hydrophone at $6\pm0.5~\mathrm{mm}$ from the explosion along the transmitted shock direction, is shown in Fig.~\ref{fig:reverberation_sensor}~(a), with its power spectral density (PSD) shown in Fig.~\ref{fig:reverberation_sensor}~(b).

Four main density peaks are identified in the sensor signal, with an average separation period $\Delta t_r$. The density peak amplitudes progressively decrease, indicating that the pressure wave reverberating inside the solid steadily weakens. In contrast, the density immediately ahead of each front increases, indicating that the water has been progressively compressed by the previously transmitted fronts. Only three reverberation cycles can be measured reliably from the sensor signal. The additional cycles visible in Fig.~\ref{fig:reverberation_video}~(d) may be too weak to produce measurable peaks and may lose their two-dimensionality while propagating in the highly disturbed water, leading to low signal-to-noise ratios. Approximating the internal wave speed by the longitudinal sound speed of Al gives

\begin{equation}\label{freqEq}
    \Delta t_r = \frac{2d}{a_0} = \frac{1}{f_r},
\end{equation}

\noindent where $d$ is the sample thickness and $f_r$ is the reverberation frequency, giving $f_r \sim 2.07~\mathrm{MHz}$ for $\Delta t_r = 484~\mathrm{ns}$. This frequency and its harmonics align with the peaks in the PSD of the density signal. Solving Eq.~\eqref{freqEq} for $a_0$, the internal wave speed is estimated as $\sim 6200~\mathrm{m/s}$, in good agreement with the longitudinal sound speed of Al listed in Table~\ref{tab:material_acoustic_properties}. This approach is similar to the time-of-flight principle used in ultrasonic pulse-echo measurements of sound speed~\cite{krautkramer1983pulse,selfridge1985approximate,anderson1998direct}, but differs in that the propagating disturbance in this experiment is a weak shock rather than a linear acoustic pulse, so the inferred value should be interpreted as an effective propagation speed.

%% Conclusions
\section{Conclusion}

This work experimentally investigated shock wave transmission generated by UEFEs through materials with different acoustic impedances and thicknesses. Free-standing explosions of $10$ and $18~\mathrm{\mu m}$ Cu foils were first characterized using high-speed shadowgraphy at a rate of $5~\mathrm{Mfps}$ and simplified one-dimensional magnetohydrodynamic simulations. The shadowgraph images showed a weak precursor compression associated with the solid-to-liquid phase transition of the foil, followed by a stronger main compression generated during the liquid-to-gas transition. The simulations reproduced the measured shock trajectories, captured the shock formation process, and provided insight into the thermodynamic, acoustic, and electrical conditions driving the explosion. Additional shadowgraph experiments recording the explosion from the side indicated that shocks remain sufficiently planar over the region of interest for both foil thicknesses.

Fiber optic hydrophone measurements of shocks generated by free-standing foil explosions were not possible because continuous emission from the expanding Cu plasma contaminated the sensor signal. To overcome this limitation, the foils were attached to a $0.1~\mathrm{mm}$-thick matte black-painted Mylar sheet, which electrically insulated the target materials from the foil and prevented plasma light emission from reaching the sensor. This enabled pressure measurements of UEFE-generated shocks after transmission through solid materials. Pressure was recovered from the measured density using several water EOS, showing significant sensitivity of the pressure retrieval to the EOS choice. The Mylar sheet only moderately attenuated the transmitted shock, with the measured transmission exceeding that predicted by the transmission coefficient of an isolated Mylar--water interface, possibly because the expanding Cu plasma sustained a high pressure on the Mylar after shock formation.

Shock transmission was then investigated through PLA, Al, and Cu target samples with thicknesses of $1.5$ and $5.0~\mathrm{mm}$. Increasing the target thickness consistently reduced the transmitted shock strength through greater attenuation over the longer propagation path within the solid. For a given thickness, the transmitted shock strength decreased with increasing acoustic impedance mismatch as a smaller fraction of the incident shock was transmitted into the water. Accordingly, PLA produced the strongest transmitted shocks, consistent with its acoustic impedance being closest to that of water, whereas Al and Cu yielded progressively weaker transmitted shocks. Additional experiments with thinner Al targets indicated that attenuation also depends on the mechanical response of the target. Samples of thickness $\leq0.5~\mathrm{mm}$ deformed and perforated, allowing the expanding Cu plasma to expand more freely and sustain the transmitted disturbance, whereas thicker samples ($\geq1.0~\mathrm{mm}$) maintained sufficient structural integrity to confine the plasma and attenuate shocks more effectively. Reverberating shocks were observed in Al samples, with measured reverberation frequencies in good agreement with the expected wave transit time inside the material.

Overall, the combination of direct pressure measurements and magnetohydrodynamic simulations provides a quantitative description of both shock transmission and the underlying foil explosion dynamics in UEFEs. The results show that thin 3D-printed PLA materials can be used as structural supports without significantly altering the transmitted shock strength, while increasing acoustic impedance mismatch and target thickness provide controlled shock attenuation through propagation losses in the solid and reduced transmission at the material--water interface. More broadly, the total transmission coefficient provides a simple and consistent parameter for relating the properties of the target material to the resulting transmitted shock strength. These findings provide quantitative guidance for selecting structural support materials in future UEFE-based experiments, enabling simplified manufacturing of complex foil geometries while preserving shock strength and planarity. While this work has focused on shock transmission, future work should investigate reflected shocks and extend the magnetohydrodynamic simulations to a multiphase framework incorporating the solid materials.

\begin{acknowledgments}
This work is supported by ERC grant DYNPRESS, project number 101042878. Funded by the European Union. Views and opinions expressed are however those of the authors only and do not necessarily reflect those of the European Union or the European Research Council Executive Agency. Neither the European Union nor the granting authority can be held responsible for them.
\end{acknowledgments}

\section*{Data Availability}
%The data that support the findings of this study are openly available in [to be filled in].
The data supporting the findings of this study will be made publicly available with the associated peer-reviewed publication.

\appendix

\section{Electrical waveforms}\label{Sec:Electrical}
\begin{figure*}[h]
    \centering
    \includegraphics[width=.93\linewidth]{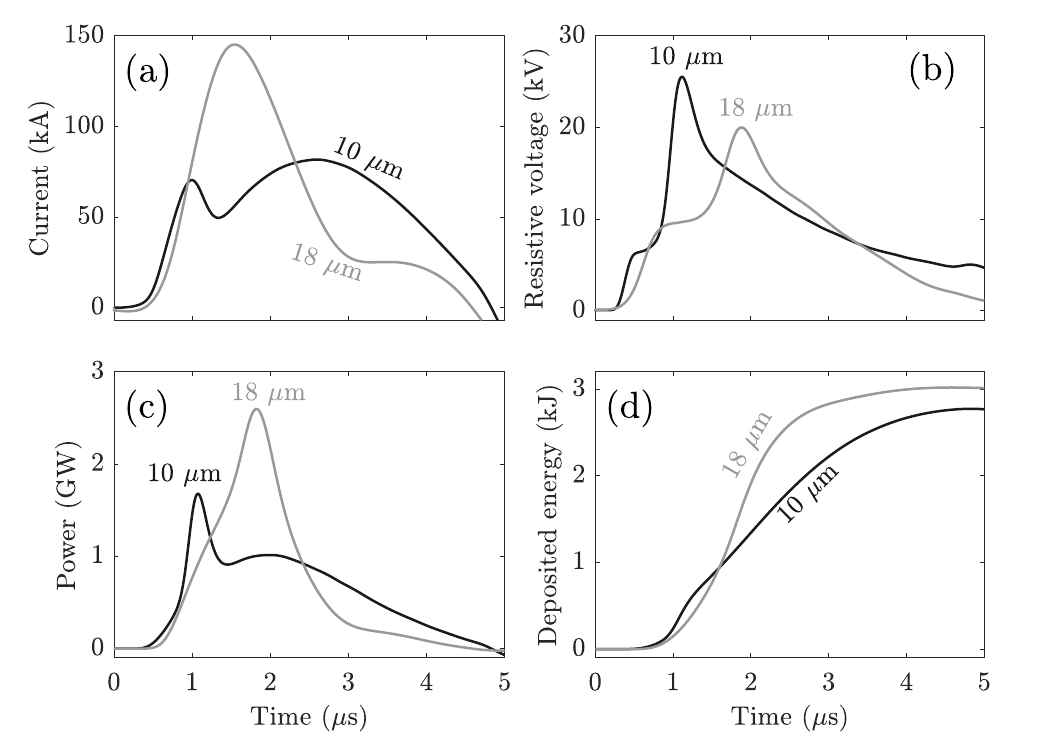}
    \caption{Electrical waveforms of the discharge current (a), resistive voltage (b), power (c), and deposited energy (d) for free-standing $10$ and $18~\mathrm{\mu m}$ Cu foil explosions. Time is relative to the current initiation.}
    \label{fig:electrical_waveforms}
\end{figure*}

The electrical waveforms corresponding to the free-standing UEFEs are included in Fig.~\ref{fig:electrical_waveforms}, including the current, resistive voltage, power, and deposited energy of the discharge. Small shot-to-shot variations were seen in the explosions on different target materials, which do not affect the present analysis. The current rise times to the first peak are $\sim1.00~\mathrm{\mu s}$ and $\sim1.45~\mathrm{\mu s}$, corresponding to currents of $\sim 70~\mathrm{kA}$ and $\sim 145~\mathrm{kA}$ for the $10~\mathrm{\mu m}$ and $18~\mathrm{\mu m}$ foils, respectively. Current restrike is observed for the $10~\mathrm{\mu m}$ foil, indicating that it is more undermassed than the $18~\mathrm{\mu m}$ case. A physically motivated definition of the explosion time corresponds to the instant of the first current peak, at which the foil material transitions from liquid to gas and further into weakly ionized plasma, while the magnetic pressure becomes smaller than the internal pressure and initiates rapid expansion of the foil \cite{hernandez2025characterization}. These times are consistent with the simulated liquid-to-gas transition described in Sec.~\ref{Sec:FoilProperties}, within an error of $\sim 0.1~\mathrm{\mu s}$.

The resistive voltages peak at $\sim 1.10~\mathrm{\mu s}$ and $\sim 1.85~\mathrm{\mu s}$, yielding $\sim 25.5~\mathrm{kV}$ and $\sim20.0~\mathrm{kV}$ for the $10~\mathrm{\mu m}$ and $18~\mathrm{\mu m}$ foils, respectively. The foil electrical inductance is computed using Eq.~\eqref{Eqimpedance}, giving an initial value of $\sim 19~\mathrm{nH}$ for both foil thicknesses. For explosions on Al and Cu samples, occasional sparks may appear after the resistive voltage peak, likely due to sample manufacturing quality and the installation of the insulating Mylar sheet, with no measurable effect on either the Mach number or density of the generated shocks. The power, computed as the product of current and resistive voltage, peaks at $\sim 1.7~\mathrm{GW}$ and $\sim 2.6~\mathrm{GW}$ for the $10~\mathrm{\mu m}$ and $18~\mathrm{\mu m}$ foils, respectively. The power peak is located much closer in time to the peak in resistive voltage than to the peak in current~\cite{oreshkin2020integral}. The deposited energy shows a monotonically increasing trend, with final energies approaching $\sim 2.75 ~\mathrm{kJ}$ and $\sim 3.00 ~\mathrm{kJ}$ asymptotically for the $10~\mathrm{\mu m}$ and $18~\mathrm{\mu m}$ foils, respectively. These values compare well with the initial available energy in the capacitors of $\sim 2.65 ~\mathrm{kJ}$, with the discrepancy attributed to the measurement diagnostics and to the simplifying assumptions of the inductance calculation, especially toward the plasma phase.

\section{Water Hugoniot curves} \label{Sec:HugoniotCurves}

This section describes the procedure used to construct the water Hugoniot curves from the selected EOS. Let subscripts 0 and 1 denote the upstream (pre-shock) and downstream (post-shock) states, respectively. The specific volume, temperature, pressure, specific internal energy, density, shock velocity, upstream sound speed, and post-shock particle velocity are denoted by $v$, $T$, $p$, $e$, $\rho$, $U_s$, $a_0$, and $u_p$, respectively. The shock Mach number $M$ and specific volumes are defined as

\begin{align}
    M &= \frac{U_s}{a_0}, \\
    v_i &= \frac{1}{\rho_i}, \qquad i=0,1 .
\end{align}

\noindent The pressure and specific internal energy are given by the selected EOS,

\begin{align}
    p_i &= p(v_i,T_i), \\
    e_i &= e(v_i,T_i), \qquad i=0,1 ,
    \label{eq:eos_closure}
\end{align}

\noindent where SESAME 7153, SESAME 7154~\cite{lyon1978sesame}, and IAPWS-95~\cite{wagner2002iapws} EOS for water are considered. The shock upstream state is fixed at

\begin{equation}
\begin{gathered}
\begin{aligned}
    \rho_0 &= 998~\mathrm{kg/m^3}, \qquad
    T_0 &= 293~\mathrm{K}, \qquad
    a_0 &= 1500~\mathrm{m/s},
\end{aligned}
\\
\begin{aligned}
    p_0 &= p(v_0,T_0), \qquad
    e_0 &= e(v_0,T_0).
\end{aligned}
\end{gathered}
\end{equation}

For each value of $M$, the downstream specific volume $v_1$ and temperature $T_1$ are iterated until the momentum jump condition and the Hugoniot energy equation~\cite{anderson2003modern} are simultaneously satisfied, respectively,

\begin{align}
    p(v_1,T_1) &= p_0 + \rho_0^2 U_s^2 \left( v_0 - v_1 \right),
    \label{eq:momentum_jump} \\
    e(v_1,T_1) - e_0 &=
    \frac{1}{2}
    \left[ p(v_1,T_1) + p_0 \right]
    \left( v_0 - v_1 \right).
    \label{eq:hugoniot_energy}
\end{align}

The post-shock pressure and specific internal energy are then recovered from the EOS, and the post-shock particle velocity is obtained from mass conservation as

\begin{equation}
    u_p = U_s \left( 1 - \frac{v_1}{v_0} \right).
    \label{eq:particle_velocity}
\end{equation}

\begin{figure*}[h]
    \centering
    \includegraphics[width=1\linewidth]{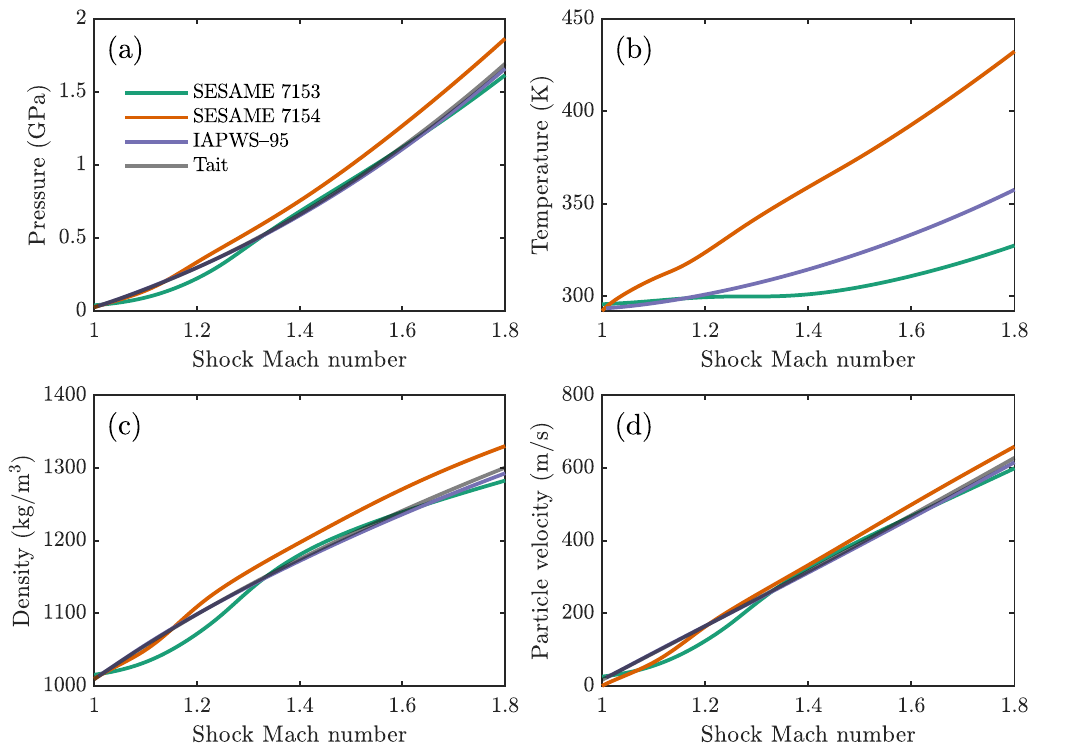}
    \caption{Water Hugoniot curves obtained using the SESAME 7153, SESAME 7154, IAPWS-95, and Tait equations of state. The panels show the post-shock state as a function of shock Mach number: (a) pressure, (b) temperature (excluding Tait), (c) density, and (d) particle velocity.}
    \label{fig:Hugoniot_curves}
\end{figure*}

For comparison, an isothermal Tait EOS~\cite{ridah1988shock} is also included for the pressure--density relation,

\begin{equation}
    p + B = (p_0 + B)\left(\frac{\rho}{\rho_0}\right)^\gamma ,
    \label{eq:tait_eos}
\end{equation}

\noindent where $B = 300~\mathrm{MPa}$ and $\gamma = 7.15$. For each prescribed $M$, Eq.~\eqref{eq:tait_eos} was combined with the momentum jump condition, Eq.~\eqref{eq:momentum_jump}, to determine the corresponding downstream density $\rho_1$ and pressure $p_1$. Since this form does not provide the specific internal energy or temperature, it was used only to compare the pressure, density, and particle velocity curves. The resulting Hugoniot curves are given in Fig.~\ref{fig:Hugoniot_curves}.

%\nocite{*}
\bibliography{references}

\end{document}